\documentclass[%
 reprint,
superscriptaddress,
 amsmath,
 amssymb,
 aps,
 prl,
 showpacs, 
 floatfix,
]{revtex4-2}

\usepackage{graphicx}
\usepackage{dcolumn}
\usepackage{bm}

\begin{document}

\preprint{APS/123-QED}

\title{Variational Scarring in Graphene Quantum Dots}

\author{J.~Keski-Rahkonen\textsuperscript{†}}
\affiliation{Department of Physics, Harvard University, Cambridge, Massachusetts 02138, USA}
\affiliation{Department of Chemistry and Chemical Biology, Harvard University, Cambridge, Massachusetts 02138, USA}
\thanks{\textsuperscript{†} These authors contributed equally to this work.}

\author{C.~Zou\textsuperscript{†}}
\affiliation{Harvard College, Harvard University, Cambridge, Massachusetts 02138, USA}
\thanks{\textsuperscript{†} These authors contributed equally to this work.}

\author{A.M.~Graf}
\affiliation{Department of Physics, Harvard University, Cambridge, Massachusetts 02138, USA}
\affiliation{Department of Chemistry and Chemical Biology, Harvard University, Cambridge, Massachusetts 02138, USA}
\affiliation{Harvard John A. Paulson School of Engineering and Applied Sciences,
Harvard, Cambridge, Massachusetts 02138, USA}

\author{Q.~Yao}
\affiliation{Department of Physics, University of California, Santa Cruz, California 95064, USA}

\author{T.~Zhu}
\affiliation{Department of Physics, University of California, Santa Cruz, California 95064, USA}

\author{J.~Velasco, Jr}
\affiliation{Department of Physics, University of California, Santa Cruz, California 95064, USA}

\author{E.J.~Heller}
\affiliation{Department of Physics, Harvard University, Cambridge, Massachusetts 02138, USA}
\affiliation{Department of Chemistry and Chemical Biology, Harvard University, Cambridge, Massachusetts 02138, USA}


\date{\today}

\begin{abstract}
A quantum eigenstate of a classically chaotic system is referred as \emph{scarred} by an unstable periodic orbit if its probability density is concentrated in the vicinity of that orbit. Recently, a new class of scarring - variational scarring - was discovered in numerical studies of disordered quantum dots, arising from near-degeneracies in the quantum spectrum associated with classical resonances of the unperturbed system. Despite the increasing body of theoretical evidence on variational scarring, its experimental observation has remained out of reach. Motivated by this dearth, we argue and demonstrate that variational scarring can occur in an elliptical quantum dot fabricated on monolayer graphene, and locally perturbed by a nanotip. Then, we further show that the fingerprint of these variational scars can potentially be detected via scanning tunneling microscopy, thus offering an attractive experimental pathway for the first validation of this puzzling quantum-chaotic phenomenon.

\end{abstract}

                              
\maketitle
\noindent
Most systems in nature are chaotic~\cite{Strogatz_book}, and the conundrum of chaos in a quantum system has long been a topic of intense study~\cite{Gutzwiller_book, Stockmann_book, Haake_book, Heller_book_2008}, dating
back to the dawn of quantum mechanics~\cite{einstein_verh.dtsch.phys.ges_19_82_1917, stone_phys.today_58_37_2005}. Although the subject of quantum chaos~\cite{berry_phys.scripta_40_335_1989} has not yet reached its final form, it already encompasses a diverse set of original phenomena from the point of view of classical chaos, such as the scarring of the eigenstates of single particles~\cite{Heller_phys.rev.lett_53_1515_1984, kaplan_ann.phys_264_171_1998,Kaplan_nonlinearity_12_R1_1999}.  

In quantum scarring, the probability density of a quantum state is imprinted with a short, unstable periodic orbit (PO) of a chaotic classical system due to quantum interference. These scars are surprising as there is no such classical concentration of density near a PO in the long time limit for a bounded chaotic system, hence standing out against the corresponding, monotonous backdrop of quantum state randomness across most of phase space. Since the initial prediction of the phenomenon by Heller~\cite{Heller_phys.rev.lett_53_1515_1984}, extensive experimental attempts have been carried out to image scarred states. Even though analogs of scars have been observed in various classical wave experiments, such as in microwave resonators~\cite{Sridhaar_phys.rev.lett_67_785_1991, Stein_phys.rev.lett_68_2867_1992}, acoustic cavities~\cite{chinnery_phys.rev.e_53_272_1996}, and fluid surface waves~\cite{kudrolli_phys.rev.e_63_026208_2001}, their direct visualization in a genuine quantum system remained elusive until the recent breakthrough ~\cite{Ge_Nature_2024} in a stadium-shaped, graphene-based quantum dot (QD).

\begin{figure}[h!]
    \centering
    \includegraphics[width=1.0\linewidth]{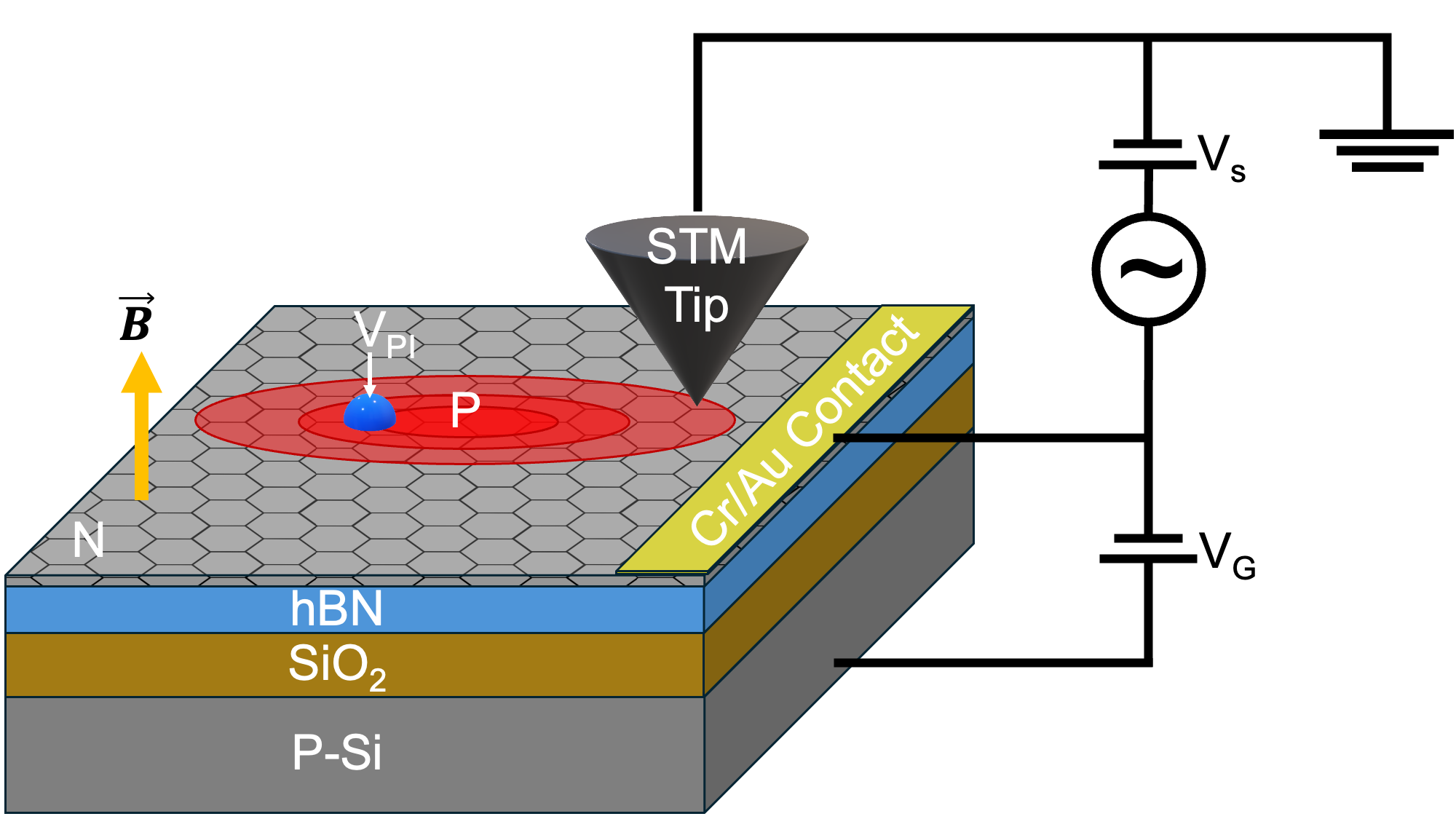}
    \caption{The figure presents the proposed scanning tunneling microscopy scheme designed for observing variational scars, in a similar manner in Ref.~\cite{Ge_Nature_2024}. It features a QD defined by an elliptical p-n junction on a monolayer of graphene (red zone), which is placed on hexagonal boron nitride (hBN). The Fermi level of the QD is controlled by a gate bias $V_g$, applied via an underlying doped silicon (P-Si) wafer separated by aFormation, prevalence, and stability of bouncing-ball quantum scars thin layer of silicon oxide (Si$\textrm{O}_2$) from the QD structure. The quantum dot can also be exposed to a constant magnetic field $\mathbf{B}$. Moreover, a nanotip (or an impurity atom) is utilized to introduce a localized disturbance within the confining potential $V_{\textrm{PI}}$ (blue dot), creating a controlled, spatially restricted perturbation within the QD, which is a necessary ingredient for variational scars. A scanning tunneling microscope tip is subsequently employed to map out the QD wavefunction at a specific probing bias $V_s$. Later, in Fig.~\ref{Fig:STM_map}, we demonstrate that this type of measurement can visually unveil the effect of variational scars taking place in the propounded QD setup.}
    \label{Fig:QD_device}
\end{figure}
 
Inspired by this progress, we suggest an experimental pathway to directly visualize a different recently discovered member of the scar family, namely variational scars, first reported in disordered two-dimensional nanostructures~\cite{Luukko_sci.rep_6_37656_2016}. Although these scars resemble the conventional scarring by Heller~\cite{Heller_phys.rev.lett_53_1515_1984}, they stem from a fundamentally different mechanism~\cite{Luukko_sci.rep_6_37656_2016, keski-rahkonen_phys.rev.lett_123_214101_2019}: the (near) symmetry of the unperturbed system leads to scarred states favored by the variational principle in the presence of a localized perturbation. As a result, the scars in the \emph{perturbed} quantum system form around the POs of the \emph{unperturbed} classical counterpart, so we call them perturbation-induced, variational scars. This scarring mechanism is ubiquitous~\cite{Luukko_sci.rep_6_37656_2016, keski-rahkonen_phys.rev.lett_123_214101_2019, keski-rahkonen_j.phys.conden.matter_31_105301_2019, keski-rahkonen_phys.rev.b_97_094204_2017, luukko_phys.rev.lett_119_203001_2017, Selinummi_Phys.Rev.B_110_235420_2024, Zhang_discover.nano_19_72_2024, phantom_scars}, but experimental validation of variational scarring has yet to be achieved. To address this issue, we investigate the potential of a realizable quantum dot on monolayer graphene to foster variational scars, allowing a scanning tunneling microscopy (STM) detection of these scarred states, in the same spirit as in Ref.~\cite{Ge_Nature_2024}.

We propose the following paradigmatic QD setup, schematically depicted in Fig.~\ref{Fig:QD_device}. The elliptical electrostatic confinement of the QD is achieved using a p-n junction within a heterostructure of monolayer graphene interfaced with sheets of hexagonal boron nitride that rests on a wafer composed of doped silicon covered with silicon oxide. A gate voltage $V_g$ can be applied to the underlying doped silicon substrate to tune the Fermi level $\varepsilon_F$ of the monolayer graphene. For imaging the shape of a quantum state, a grounded STM tip can be utilized, with a probing voltage bias $V_s$ applied via the graphene layer. Moreover, a second tip can generate a local perturbation, creating a potential bump or dip depending on the applied voltage, within the QD confinement. This proposed QD device aligns not only with Ref.~\cite{Ge_Nature_2024} but also with other recent advancements on quantum state control and imaging~\cite{Ge_nano.lett_20_8682_2020, Ge_nano.lett_21_8993_2021}. We attribute the alluring opportunity of visualizing scars in the QD device to two pivotal developments: higher Q-factor compared to foregoing experiments, such as the successfully imaging of a quantum corral~\cite{Heller_Nature_369_464_1994}, and an exquisite improvement in energy resolution up to the level of individual eigenstates, as shown in Refs.~\cite{Ge_nano.lett_20_8682_2020, Ge_nano.lett_21_8993_2021}.    

\begin{figure}[h!]
\centering
\includegraphics[width=1.00\linewidth]{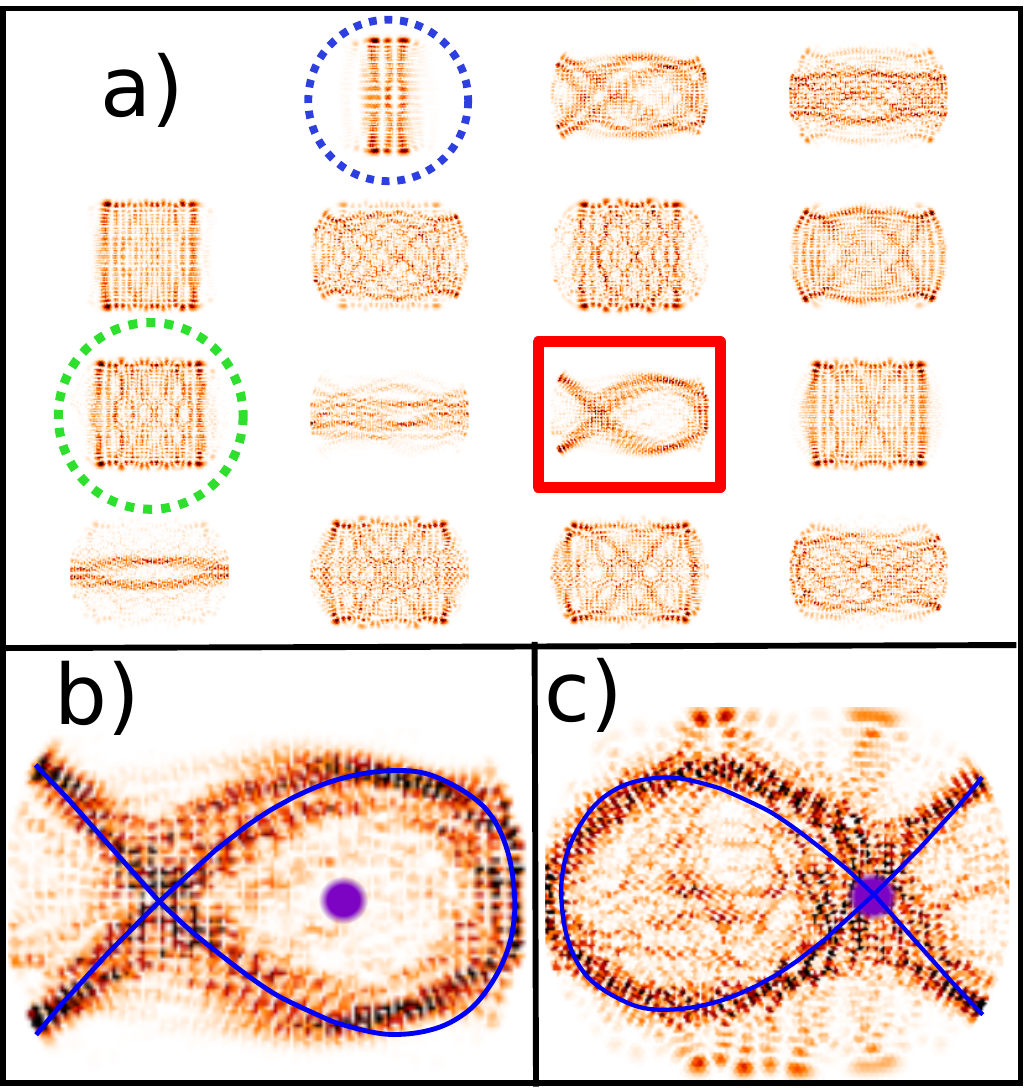}
\caption{The figure summarizes the observed features of the variational scars in the elliptical QD perturbed by a nanotip. Subplot (a) displays a collection of high-energy eigenstates of the Hamiltonian described by Eq.~\ref{Eq_Hamiltonian}, with some strongly scarred by classical Lissajous periodic orbits. Subplot (b) highlights an exemplary scar [marked by the red square in (a)], where the probability density concentrates along the Lissajous orbit in the shape of the symbol alpha (blue curve). The scar is also oriented in such a way that its loop avoids the bump (violet dot), effectively minimizing the perturbation caused by the nanotip. These variational scars are accompanied by companion scars that, as shown in subplot (c), maximize the the effect of perturbation, rather than minimize it. Some of the displayed quantum states exhibit classical bouncing-ball-like motion (blue dashed circle), while others remain almost unaffected by the nanotip (green dashed circle).}
\label{Fig:Scar_examples}
\end{figure}

To model the device outlined above, we turn to a perturbed two-dimensional QD with the following tight-binding Hamiltonian:
\begin{equation}\label{Eq_Hamiltonian}
\begin{split}
    \mathcal{H} &= -t\sum_{\langle i,j\rangle} \left(  a_i^{\dagger}b_j + b_j^{\dagger}a_i\right)\\ &+ \sum_i \Big[ V_{\textrm{ext}} + V_{\textrm{tip}} + V_{\textrm{imp}} \Big] \left(a_i^{\dagger}a_i + b_i^{\dagger}b_i \right),
\end{split}
\end{equation}
where $a$ ($a^{\dagger}$) and $b$ ($b^{\dagger}$) are the annihilation (creation) operators on the graphene sublattices A and B, respectively. We consider the conventional nearest-neighbor coupling;  For simplicity, we set the hopping energy to be $t= 1.0$.~\footnote{We want to emphasize that this convention measures energies in units of $2.7 \, \textrm{eV}$, and then the distances are given in lattice spacings of $d = 0.142 \, \textrm{nm}$. } Moreover, unless otherwise stated, we assume the noise term $V_{\textrm{imp}} = 0$. In fact, our simulation parameters closely match those of the experimental setup described in Ref.~\cite{Ge_Nature_2024}; however, we want to stress that the following considerations extend beyond these specific parameter choices~\footnote{Here, the graphene hopping $t= 1$, potential parameters $\omega_0 =2$ and $\alpha = 0.5$, and the nanotip parameters $A= 0.35$ and $\sigma=1.8$}.

Previous experimental and theoretical studies have generally validated the use of the harmonic approximation to represent the external confining potential of electrons in a QD~\cite{reimann_rev.mod.phys_74_1283_2002}, even extending to the quantum Hall regime under a strong magnetic field~\cite{rasanen_phys.rev.B_77_041302_2008, rogge_phys.rev.lett_105_046802_2010}. Therefore, the confinement potential is assumed to be an elliptical harmonic oscillator, expressed as
\begin{equation}
    V_{\textrm{ext}}(\mathbf{r}) = \frac{1}{2}\omega_0\left( \alpha^2 x^2  +  y^2\right),
\end{equation}
where $\omega_0$ is the overall confinement strength and $\alpha$ represents the ellipticity. We set $\omega_0 = 2$, aligning with the experimental confinement strengths reported for similar QDs. Without loss of generality, we focus on a special value of $\alpha = 0.5$ where, crucially,  accidental degeneracies occur in the unperturbed system associated with the Lissajous orbits existing in the classical counterpart. We note that Lissajous scars appear at various values of ellipticity $\alpha$ (see Ref.~\cite{keski-rahkonen_phys.rev.lett_123_214101_2019}). Moreover, as will be shown later, the ellipticity $\alpha$ does not need to be precisely matched to the classical resonant conditions of the Lissajous POs. Finally, we emphasize that the experimental feasibility of this type of graphene-based quantum dot has been substantiated by recent works, such as  Refs.~\cite{Ge_Nature_2024, Ge_nano.lett_21_8993_2021, Ge_nano.lett_20_8682_2020}.

A Gaussian potential bump or dip at location $\mathbf{r}_0$ has been shown to be an ideal model for a local perturbation induced by a nanotip~\cite{blasi_phys.rev.B_87_241303_2013, boyd_nanotechnology_22_185201_2011,bleszynski_nano.lett_7_2559_2007}:
\begin{equation}
    V_{\textrm{tip}}(\mathbf{r}) = A_{\textrm{tip}} \textrm{exp} \left[ \frac{\vert \mathbf{r} - \mathbf{r}_0 \vert^2}{\sigma_{\textrm{tip}}^2} \right] 
\end{equation}
described by its amplitude $A = 0.35$ and width $\sigma = 1.8$. Similarly, randomly distributed Gaussian bumps are also employed to model impurity noise in a QD~\cite{hirose_phys.rev.B_65_193305_2002, hirose_phys.Rev.B_63_075301_2001, guclu_phys.rev.B_68_035304_2003}. Here, we have selected the values of the amplitude and width so that the nanotip induces strong scarring in the relevant energy regime. Although we spotlight the local perturbation caused by a nanotip in this study, a similar effect can be engineered by introducing an adatom(s) at a specific location(s) on the graphene layer, or by leaving an n-doped region within the p-doped domain that forms the QD during the manufacturing phase. These alternatives avoid the difficulty of arranging an STM tip in conjunction with the nanotip that generates the perturbation $V_{\textrm{tip}}$. Nonetheless, a multitip approach would provide a new level of controllability (see, e.g. Refs.~\cite{Voigtlander_Rev.Sci.Instrum_89_101101_2018, Leis_rev.sci.instrum_93_013702_2022, Hus_phys.rev.lett_119_137202_2017}).

 In Fig.~\ref{Fig:QD_device}, we have included the possibility of a homogeneous magnetic field that is perpendicular to the QD. The previous studies,~\cite{keski-rahkonen_phys.rev.lett_123_214101_2019,keski-rahkonen_j.phys.conden.matter_31_105301_2019, keski-rahkonen_phys.rev.b_97_094204_2017} variational scars emerged in a disordered QD when the effective potential stemming from the external confinement and magnetic field satisfies the classical resonant condition for the Lissajous orbits. Alternatively, a form of variational scarring can arise when spin-orbit coupling replaces local perturbation in a QD~\cite{Zhang_discover.nano_19_72_2024, phantom_scars}. These systems also appear suitable for scar detection through STM~\cite{Ge_nano.lett_20_8682_2020, Ge_nano.lett_21_8993_2021, Velasco_nano.lett_18_5104_2018, Freitag_nat.nanotechnol_13_392_2018}, and for identifying indirect scar signatures via NMR~\cite{Kuzma_science_281_686_1998, Barrett_phys.rev.lett_74_5112_1995} and spin-dependent transport measurements~\cite{Berger_nanomaterials_11_1258_2021, Sun_appl.phys.lett_117_052403_2020, Peng_appl.phys.lett_118_082402_2021, Wang_appl.phys.lett_118_083105_2021} . However, we will focus on the scenario without the external magnetic field, which still offers the potential for direct visualization of variational scars, as demonstrated below.

We compute thousands of eigenstates of our Hamiltonian, utilizing the eigenfunction solver in \texttt{Kwant}~\cite{Kwant}. As expected, based on the theory of varitional scarring,~\cite{keski-rahkonen_phys.rev.lett_123_214101_2019, keski-rahkonen_phys.rev.b_97_094204_2017},  multiple high-energy eigenstates of the elliptical QD show strong scarring when perturbed by the nanotip, characterized by Lissajous orbits of the unperturbed classical system. In Fig.~\ref{Fig:Scar_examples} (b) and (c), we show a collection of probability densities of the solved eigenstates, spotlighting two archetypal scars along with its classical PO and the bump in Fig.~\ref{Fig:Scar_examples} (b) and (c). Moreover, the collection in Fig.~\ref{Fig:Scar_examples} (a) includes quantum states with classical "bouncing-ball-like" motion (blue dashed circle) as well as so-called remnants (green dashed circle), which appear to be almost unaffected by the applied perturbation.

\begin{figure}[h!]
\centering
\includegraphics[width=1\linewidth]{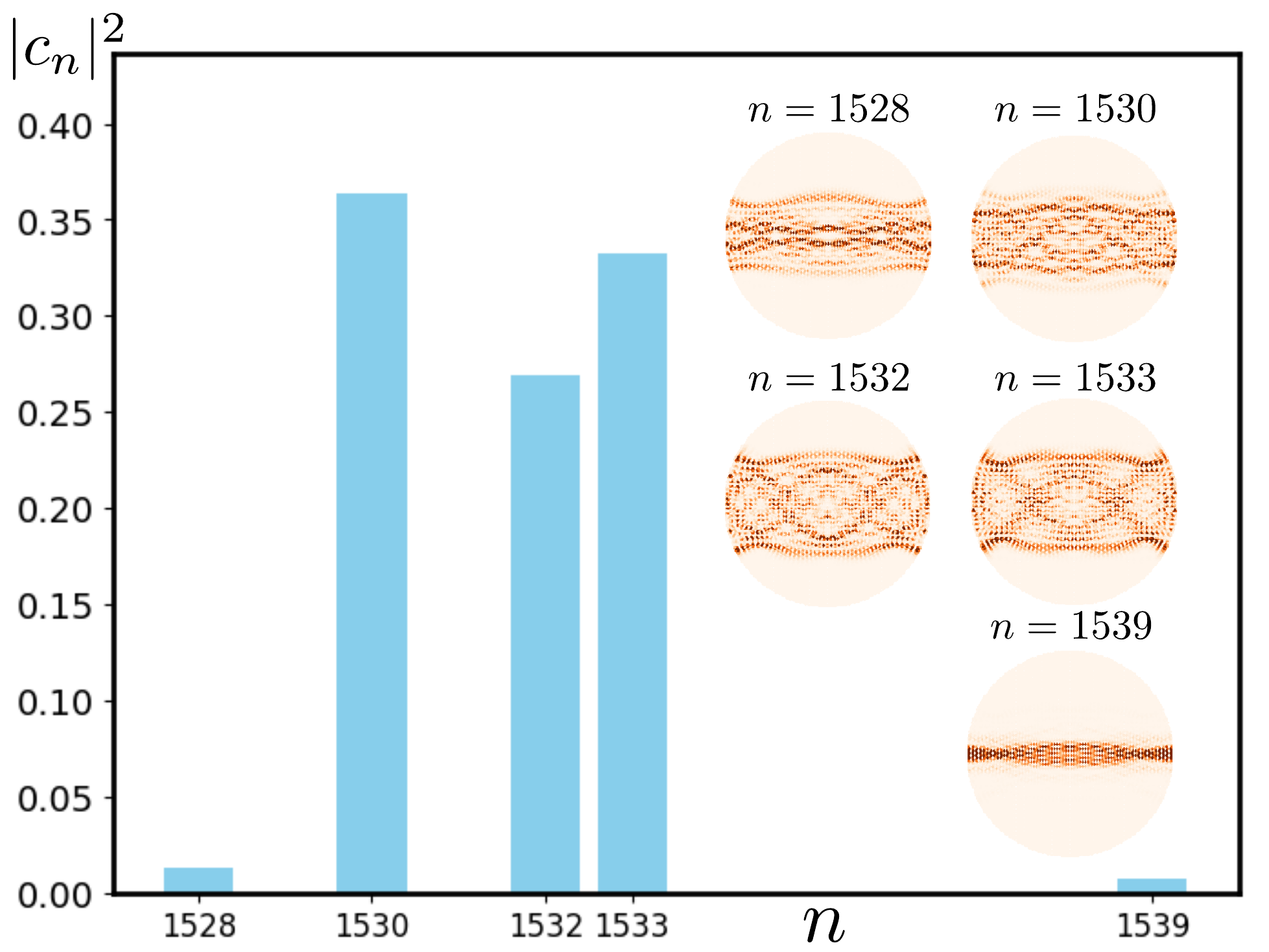}
\caption{Subspectrum analysis: the scarred state $\vert \textrm{Scar} \rangle$ presented in Fig. 1(b) is expanded in the basis of unperturbed states $\vert \phi_n \rangle$, with the corresponding coefficients $\vert c_n \vert^2 = \vert \langle \textrm{Scar} \vert \phi_n \rangle \vert^2$ that reveals that the representative scar arises from a relatively small subset of the near-degenerate eigenstates of the system without the nanotip.}
\label{Fig:Scar_decomposition}
\end{figure}

We emphasize that this type of scarred state does not exist in the unperturbed QD, whose eigenstates $\vert \phi_n \rangle$ are instead expressible in terms of Hermitian polynomials~\footnote{The eigenstates of an elliptical are $\vert \phi_{n, m}\rangle \propto H_m(\sqrt{\alpha\omega_0}x) H_n(\sqrt{\omega_0}y)\exp[-\omega_0(\alpha^2 x^2 + y^2)/2]$, where $H_m(\cdot)$ is the
Hermite polynomial of order $m$. The corresponding energy spectrum shows degeneracies at commensurable frequencies, that is, when the ellipticity is a rational number.}. This observation strongly supports the assertion that the scars in the QD are variational in nature. To further confirm the origin of the reported scarring, we show two other trademarks unique to variational scarring. First, some of the observed scars align themselves to minimize the perturbation caused by the nanotip, as seen in Fig.~\ref{Fig:Scar_examples} (b); whereas the others counterintuivitely seek to maximize it, as displayed in Fig.~\ref{Fig:Scar_examples} (c). Second, we expand the scar presented in Fig.~\ref{Fig:Scar_examples} (b) in terms of the eigenstates of the unperturbed elliptical QD as $\vert \textrm{Scar} \rangle = \sum_n c_n \vert \phi_n \rangle$. This type of ``subspectrum'' analysis of the expansion coefficients $\vert c_n \vert^2$ reveals that the scars are formed by a small set of states close in energy, which is consistent with the theory of variational scarring~\cite{Luukko_sci.rep_6_37656_2016, keski-rahkonen_phys.rev.b_97_094204_2017,keski-rahkonen_j.phys.conden.matter_31_105301_2019, keski-rahkonen_phys.rev.lett_123_214101_2019, Selinummi_Phys.Rev.B_110_235420_2024}. For instance, the showcased Lissajous scar is primarily ($\sim 90\, \%$) conjured by three nearly degenerate eigenstates of the elliptical QD without the perturbation, as shown in Fig.~\ref{Fig:Scar_decomposition}.

\begin{figure}[h!]
\centering
\includegraphics[width=1.00\linewidth]{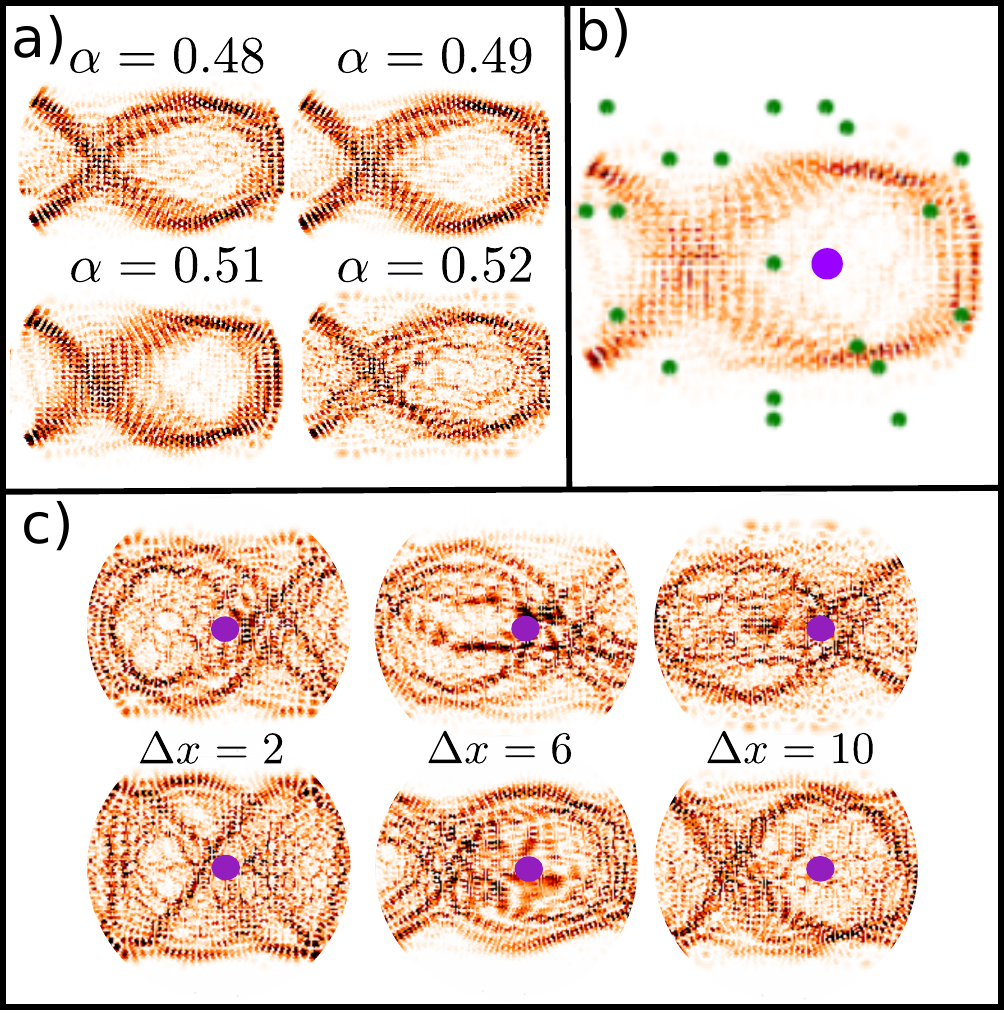}
\caption{Figure elucidates the robustness of the variational scars. Subplot (a) shows the persistence of this type of scarring near to the optimal ellipticity of $\alpha = 0.50$ against small deformations in QD confinement, when the underlying Lissajous orbits do not exist. Furthermore, subplot (b) demonstrates that the scars induced by the nanotip (violet dot) can endure under impurity noise, modeled as randomly located Gaussian bumps (green dots). On the other hand, subplot (c) reveals the sensitivity of the scar associated with a specific Lissajous orbit to the location of the nanotip, as it strives to either maximize (upper row) or minimize (lower row) the perturbation. While the strength of the scarring depends on the position of the perturbation, the alpha-shape scars are present at the optimal ellipticity $\alpha = 0.5$ across various tip positions.}
\label{Fig:Scar_robostness}
\end{figure}

Most importantly from the experimental point of view, the reported scars are robust. In Fig.~\ref{Fig:Scar_robostness} (a), we show that scars persist when the ellipticity $\alpha$ of the confinement is varied, counterintuitively even under conditions where the Lissajous POs do not exist in the unperturbed classical system, a feature explained by the variational scarring theory~\cite{keski-rahkonen_phys.rev.lett_123_214101_2019} and discussed further below. Furthermore, even when the device is affected by noise ($V_{\textrm{imp}} \neq 0$), modeled~\footnote{More explicitly, the noise has form $V_{\textrm{imp}}(\boldsymbol{r}) = A_{\textrm{imp}}\sum_i \exp \left( \vert \boldsymbol{r} - \boldsymbol{r}_i\vert^2/\sigma_{\textrm{imp}}^2 \right)$ with $A_{\textrm{imp}} = 0.15$ and $\sigma_{\textrm{imp}} = 0.15$. In the energy range of interest, we have several bumps within the classically allowed area.} as randomly scattered Gaussian bumps similar to the nanotip but smaller in amplitude (0.15) and width (0.8), some states remain distinctly scarred, as shown in Fig.~\ref{Fig:Scar_robostness} (b), which aligns with the behavior of bouncing-ball scars reported in Ref.~\cite{Selinummi_Phys.Rev.B_110_235420_2024}. This kind of noise  is able to trigger scarring on its own~\cite{keski-rahkonen_phys.rev.lett_123_214101_2019}, but here we focus on the scarring induced by the nanotip, providing an additional level of control over the generated scars.

The focus here on a single (main) perturbation also reveals another fascinating opportunity of variational scarring which has been passed over in the previous studies emphasizing multiple similar, randomly distributed impurities~\cite{keski-rahkonen_phys.rev.lett_123_214101_2019, keski-rahkonen_phys.rev.b_97_094204_2017, keski-rahkonen_j.phys.conden.matter_31_105301_2019, luukko_phys.rev.lett_119_203001_2017, Luukko_sci.rep_6_37656_2016}. Generally, strong variational quantum scars are expected near commensurable frequency ratio $\omega_x/\omega_y = p/q $, denoted as $(p,q)$. Classical POs only exist when the frequencies are commensurable, meaning $\omega_x/\omega_y$ is a rational number, with $p$ and $q$ being coprime. Geometrically, the particle returns to its initial position with the original velocity after completing $p$ and $q$ oscillations between the vertical and horizontal turning points, respectively. Conversely, incommensurable frequencies lead to quasiperiodic motion, resulting in ergodic behavior on a torus~\cite{Strogatz_book}.

On the quantum side, (near) commensurable confinement frequencies give rise to special (nearly) degenerate states, referred as a ``resonant'' set, which are related a family of classical POs. Some linear combinations within this resonant set are scarred by these POs. When a moderate perturbation is applied, it generates eigenstates that are linear combinations of a single resonant set. According to the variational theorem, this process essentially involves extremizing the perturbation within the nearly degenerate subspace of the resonant set. Consequently, scarred states are favored because the perturbation is spatially localized. Therefore, the extremal eigenstates from each resonant set often exhibit scars corresponding to the associated POs. The classical frequencies can be merely quasiresonant as long as the perturbation looms large enough to overcome the imperfect degeneracy.

In contrast to the typical scenario involving multiple localized perturbations~\cite{keski-rahkonen_j.phys.conden.matter_31_105301_2019, keski-rahkonen_phys.rev.b_97_094204_2017, keski-rahkonen_phys.rev.lett_123_214101_2019, luukko_phys.rev.lett_119_203001_2017, Luukko_sci.rep_6_37656_2016}, we can selectively target scars of different geometries corresponding to distinct classical resonances based on the position of the nanotip. For instance, instead of the $(2,3)$ resonance scar shown in Fig.~\ref{Fig:Scar_examples}(b) resembling the alpha shape, the ellipticity $\alpha = 0.5$ would typically produce scars from the $(1,2)$ resonance, manifesting as either a parabola or a figure-eight. Both resonances are energetically close to each other. More explicitly, the $(1,2)$ resonance exhibits exact degeneracy, while the $(2,3)$ resonance is only approximately degenerate for $\alpha = 0.5$. However, unlike the case of many random bumps~\cite{keski-rahkonen_phys.rev.lett_123_214101_2019}, the scarred state from the $(2,3)$ resonance offers more effective extremization of the perturbation due to the precise placement of the nanotip. Notably, the self-crossing point of the $(2,3)$ resonance provides a particular advantage in minimizing and maximizing the perturbation [see, Fig.~\ref{Fig:Scar_examples}(b) and (c), respectively]. 

On the other hand, moving the nanotip affects the level of scarring, as illustrated in Fig.~\ref{Fig:Scar_robostness} (c), for three different tip positions shifted from the center of the QD by $\Delta x = 2, 6, 10$ along the positive x axis [the ``optimized" scars shown in Figs.~\ref{Fig:Scar_examples}(b) and (c) have $\Delta x = 10$]. The scars are weaker when the bump caused by the tip does not align with a position where the (1,2) resonance scar can optimally minimize or maximize the perturbation, i.e., by coinciding with or avoiding the self-crossing point of the scar and the bump. Nevertheless, noticeable concentrations of probability density along the corresponding Lissajous orbits are observed in some of the perturbed states. Furthermore, since the potential and, consequently, the periodic orbits scale with energy, the effect of the tip position can be mitigated by either focusing on a different energy range or slightly adjusting the confining potential.

Next, taking a step towards a first experimental visualization of the phenomenon, we demonstrate the opportunity for observing these scarred states in STM measurements. For this purpose, we assess the local density of states (LDOS) at the probing bias $V_s$, which is directly linked to the differential conductance measured by a STM tip sweeping through a QD device, as
\begin{equation}\label{Eq_differential_conductivity}
    \rho(V_s, \mathbf{r}) = \frac{\eta}{\pi}\sum_n \frac{\vert \psi_n(\mathbf{r})\vert^2}{ (\varepsilon_F + eV_s + \Delta V - \varepsilon_n)^2 + \eta^2},
\end{equation}
employing the previously resolved eigenstates $\psi(\mathbf{r})$, and the energy resolution is set to be $\eta$, in line with the previous experiments of the similar fashion~\cite{Ge_nano.lett_20_8682_2020, Ge_nano.lett_21_8993_2021}. 

\begin{figure}[h!]
    \centering
    \includegraphics[width=1.\linewidth]{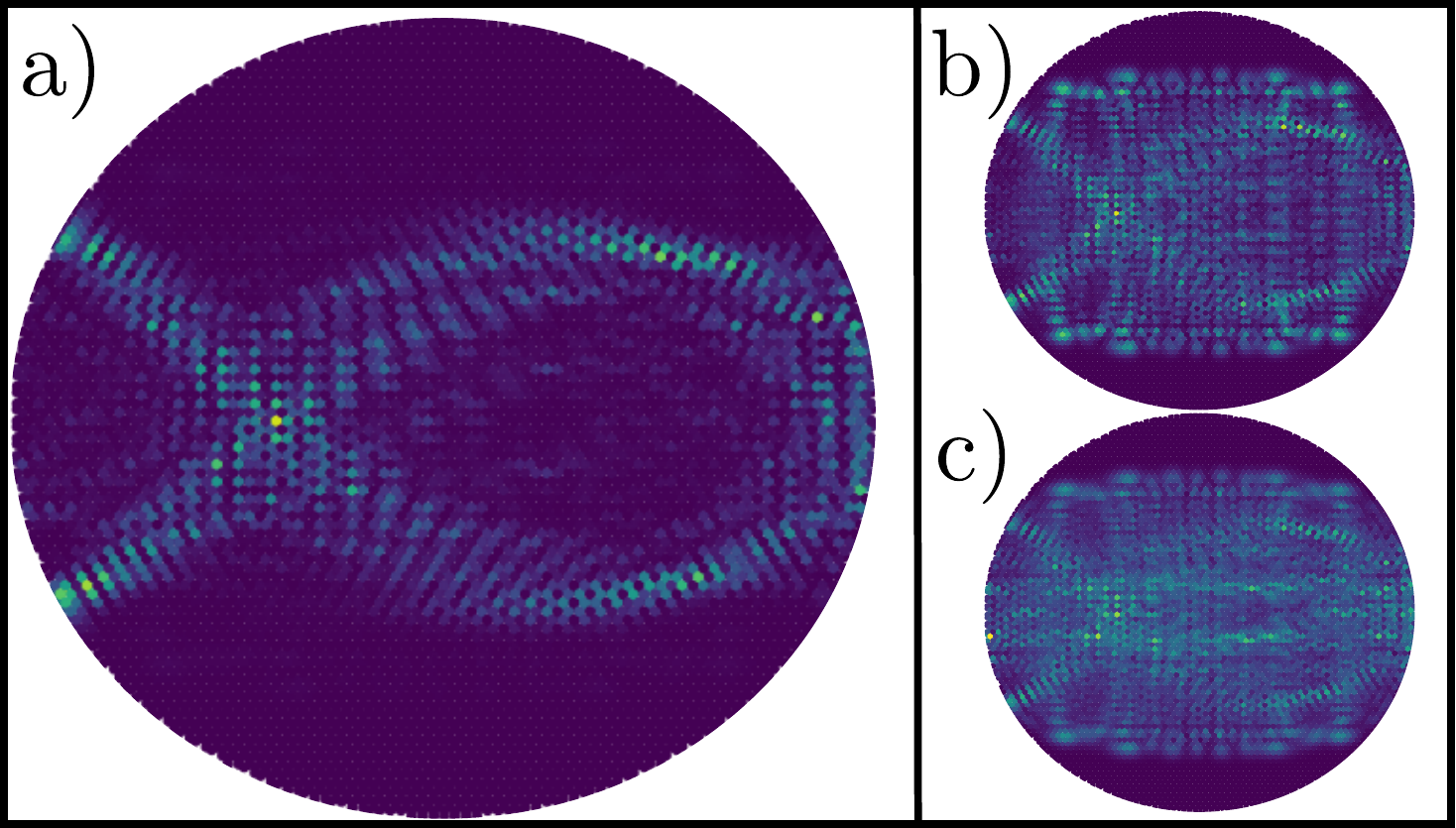}
    \caption{Figure shows the constructed LDOS map for a hypothetical differential conductance $\partial I /\partial V$ measurement. The probing bias $V_s$ of the simulated STM tip is adjusted to match with the example scar of Fig.~\ref{Fig:Scar_examples} (b). As seen in (a), the corresponding LDOS map contains a distinct imprint of the Lissajous orbit associated with the targeted scarred state when $\Delta V = 0$  and $\eta = 10^{-4}$. The scar pattern persists even when $\Delta V = 3.5\times 10^{-3}$ in (b) and $\eta = 10^{-3}$ in (c).}
    \label{Fig:STM_map}
\end{figure}

In particular, we highlight a situation where the probing bias $V_s$ is adjusted to match with the scarred state of Fig.~\ref{Fig:Scar_examples}(b). Figure~\ref{Fig:STM_map} (a) presents the corresponding spatial distribution of the LDOS that paints a discernible pattern resembling the target scar at  moderate energy resolution $\eta = 10^{-4}$ with $\Delta V  = 0$. Remarkably, the scar pattern remains apparent even when the probe voltage is mismatched, like in Fig.~\ref{Fig:STM_map}(b) with $\Delta V = 3.5 \times 10^{-4}$, or when the energy resolution is degraded to $\eta = 10^{-3}$, as shown in Fig.~\ref{Fig:STM_map}(c). Notably, the considered resolution range, $10^{-3} \le \eta \le 10^{-4}$, lies well within current experimental capabilities, as demonstrated, for example, in Ref.~\cite{Ge_Nature_2024}.

Whereas previous works, such as, Refs.~\cite{keski-rahkonen_j.phys.conden.matter_31_105301_2019, keski-rahkonen_phys.rev.b_97_094204_2017, keski-rahkonen_phys.rev.lett_123_214101_2019, Luukko_sci.rep_6_37656_2016} have primarily focused on establishing the theoretical foundation for variational scarring; our work advances this foundation towards concrete experimental realization, within a highly tunable platform such of graphene exhibiting unique features, such as pseudo-relativistic behavior~\cite{Ge_Nature_2024, Ge_nano.lett_20_8682_2020, Ge_nano.lett_21_8993_2021}.
The simple example above thus underlines the point that the variational scarring can indeed lead to a resilient observable STM signature. In other words, the STM can provide an experimentally feasible way to \emph{directly} visualize the effect of variational scars, as opposed to the prevailing approach of deducing their presence incidentally through transport measurements, such as in Refs.~\cite{narimanov_phys.rev.lett_80_49_1998, Wilkinson_nature_380_608_1996, Cabosart_nano.lett_17_1344_2017, Burke_phys.rev.lett.104.176801_2010, Zhang_Phys.Rev.B_101_085404_2020}. 
Furthermore, we note that previously considered variational scarring within a semiconductor QDs, such as studied in Refs.~\cite{keski-rahkonen_j.phys.conden.matter_31_105301_2019, keski-rahkonen_phys.rev.b_97_094204_2017, keski-rahkonen_phys.rev.lett_123_214101_2019, Selinummi_Phys.Rev.B_110_235420_2024}, is \emph{not} well-suited for an STM observation, thereby enhancing the appeal of the strategy proposed here. While optical analogs, such as the one presented in Ref.~\cite{keski-rahkonen_phys.rev.lett_123_214101_2019}, provide an alluring opportunity, the suggested STM scheme offers significant value in observing variational scarring within a \emph{genuine} quantum environment. 

In addition to facilitating the observation of variational scars, the proposed approach may be also employed to explore a necessary but subtle consequence of quantum scarring, namely antiscarring~\cite{antiscarring, Lu_antiscarring}. An interesting avenue of future research is to investigate if there exists a hybrid form of variational~\cite{Luukko_sci.rep_6_37656_2016, keski-rahkonen_phys.rev.b_97_094204_2017, keski-rahkonen_j.phys.conden.matter_31_105301_2019, keski-rahkonen_phys.rev.lett_123_214101_2019} and relativistic scars~\cite{Huang_phys.rev.lett_103_054101_2009, Xu_phys.rev.lett_110_064102_2013, Song_phys.rev.research_1_033008_2019}, or a possible transition between these two types of scarring, as the energy is varied. Another interesting possibility is to utilize the presented quantum device as a platform to study quasiscarred states in a spirit similar to that of Ref.~\cite{Lee_phys.rev.lett_93_164102} that might occur due to the roughness and atomic structure of the underlying monolayer graphene, and lattice scars within a more general framework in contrast to the previous study~\cite{Fernande_new.j.phys_16_035005_2014}.

In summary, we have demonstrated the existence of variational scars induced by a nanotip in a graphene QD, extending the validity of the scar theory. Furthermore, the effect of perturbation-induced scarring has been shown to be detectable in STM measurements, marking a significant step towards directly visualizing variational scars very first time.

\begin{acknowledgments}
We are thankful for the useful discussions with E. R{\"a}s{\"a}nen, A. Mert Bozkur, and Z. Ge. 
Q. Y., T. Z, and J.V.J., acknowledge support from the Gordon and Betty Moore Foundation award 10.37807/GBMF11569. Furthermore, J.K.-R. thanks the Oskar Huttunen Foundation for financial support. This project is also supported by the National Science Foundation (Grant No.~2403491). 
\end{acknowledgments}


\bibliography{references}

\end{document}